\documentclass[twocolumn,showpacs,preprintnumbers,amsmath,amssymb,prd]{revtex4}

\usepackage{graphicx}
\usepackage{dcolumn}
\usepackage{bm}

\begin{document}

\preprint{Desy 08-006, ZMP-HH/08-1}

\title{Stable cosmological models driven by a free quantum scalar field}

\author{Claudio Dappiaggi}
\email{claudio.dappiaggi@desy.de}
\author{Klaus Fredenhagen}
\email{klaus.fredenhagen@desy.de}
\author{Nicola Pinamonti}
\email{nicola.pinamonti@desy.de}
\affiliation{{\it II} Institut f\"ur Theoretische Physik, Universit\"at Hamburg,\\
Luruper Chaussee 149, D-22761 Hamburg, Germany.}

\date{\today}

\begin{abstract}
In the mathematically rigorous analysis of semiclassical Einstein's equations, the renormalisation of the 
stress-energy tensor plays a crucial role. We address such a topic in the case of a scalar field with both
arbitrary mass and coupling with gravity in the hypothesis that the underlying algebraic quantum state is
of Hadamard type. Particularly, if we focus on highly symmetric solutions of the semiclassical Einstein's
equations, the envisaged method displays a de Sitter type behaviour even without an a priori introduced 
cosmological constant. As a further novel result we shall show that these solutions turn out to be stable. 
\end{abstract}

\maketitle
 
\section{Introduction}

A landmark in present days observational cosmology has been set by means of the measurement of the type IA
supernovae red shift which, as a byproduct, proved that the Universe is undergoing a phase of accelerated
expansion. Such a result, also combined with the most recent data collected in several other experiments, 
suggests that, in order to explain the present state of our Universe, we must take into account
the presence of a ``dark energy'' playing the role of an effective cosmological constant. 
From a theoretical point of view we still lack a full-fledged satisfactory model for dark energy and such a 
problem
was tackled in the past in several ways, the most notables being by means either of a yet unobserved 
classical scalar field coupled to gravity \cite{Linde, Mukhanov} or of a modified theory of gravity itself
(see \cite{Nojiri} and references therein for a recent review).

In the present paper our aim is to consider the backreaction of a massive quantum scalar field coupled to gravity in 
order to discuss the role played by quantum effects in the framework of cosmological models.
The interest in backreaction effects of quantum fields in cosmology is not new since, already in the
eighties,
Starobinsky \cite{Star80} addressed the same topic taking into account a massless scalar field conformally coupled to gravity (see also \cite{Vile85}). 
The endpoint of Starobinsky seminal paper was the construction of a graceful exist from a de Sitter phase of 
rapid expansion. Using quantum property of the source fields he observed that such a de Sitter spacetime is 
an unstable solution of the semiclassical Einstein's equations (see also \cite{Ford84}). More recently, in 
\cite{Shapiro}, Shapiro and Sola also considered the massive case in a similar way. They obtained as well a 
smooth exit from an inflationary phase. Since this is a topic partly far away from our goals, we shall 
consider anew such case, namely we study the semiclassical Einstein's equation
$$
G_{ab}=8\pi G \langle T_{ab}\rangle_\omega,
$$
where the left hand side is the standard Einstein's tensor whereas the right hand side is the expectation
value for the stress-energy tensor in the state $\omega$. 
It is a well known problem that the latter gives origin to divergences.
Hence it is compulsory to invoke a renormalisation procedure and, within this perspective, we would like to carry on our analysis along the lines discussed by Wald, using the point splitting regularisation.

In a series of papers \cite{Wald77,Wald78}, Wald
sets out five axioms that need to be satisfied in order to have a renormalised stress-energy tensor that can be used
in order to have possible  meaningful  semiclassical solutions of the Einstein's equation.
Sticking to such a perspective we shall show that, in some physically motivated limits, we can find a stable
solution to the semiclassical Einstein's equation. 
This leads to a great difference from the original Starobinsky model where, on the 
opposite, an unstable behaviour is displayed. To this end, we must bear in mind the following message 
already conveyed to us in \cite{HW01,HW02}: 
the renormalisation of the stress-energy tensor suffers of some ambiguities encoded in a modification of the 
action by the addition of terms depending only on the curvature and on the parameters describing the fields
such as for example the mass.
This arbitrariness is then encoded in the renormalisation parameters present in front of this arbitrary terms.
In the forthcoming discussion we shall fix the renormalisation parameters requiring a physically meaningful
theory and invoking the principle of general local covariance \cite{BFV}.
It will also turn out that the original result due to Starobinsky in the case of conformal invariant fields 
corresponds to another choice of the renormalisation constants; hence, employing a 
different criterion, the system under analysis displays a rather physically different behaviour. 

For a more mathematically oriented reader a few more comments are in due course.
Since we are interested in solutions of the semiclassical Einstein's equation, where quantum matter acts as a 
source for the gravitational field, we need to employ a quantisation scheme independent from the spacetime 
itself. Such a conceptual problem was recently addressed in a work due to Brunetti Fredenhagen and Verch 
\cite{BFV}. They showed that it is possible to simultaneously quantise on all spacetimes and the 
quantisation scheme in this framework corresponds to assign a functor between the category of spacetimes 
(${\sf Man}$) and the category of local Algebras ${\sf Loc}$ generated fields.
Furthermore such a functor transforms covariantly under any local transformation.
Unfortunately, while also fields transform covariantly under isometries, a similar conclusion cannot be drawn
for states. Therefore, since we are interested in expectation values of fields, we are forced to select a 
class of the mentioned states enjoying some suitable physical properties and in the framework of FRW 
spacetimes, this naturally leads to select the class of the so-called adiabatic states.
Starting from these premises we are now ready to use, within this abstract scheme of analysis, quantum matter
as a source for the gravity  whereas the role of Einstein's equations will select a particular set of objects 
in ${\sf Man}$, as a sort of consistency check. To rephrase, even if we can quantise in all the spacetimes 
simultaneously, once a family of states is chosen, only in few of those spacetimes 
the semiclassical Einstein's equations hold true.

\vskip .3cm

After fixing some notation, in the next section we shall recall briefly the renormalisation procedure we 
shall employ. In the third section we shall perform a suitable choice for the quantum state and, then we will
discuss the associated solutions of the semiclassical Einstein's equations. In the fourth section we shall 
justify this hypotheses by means of physical motivations. Finally some conclusion are drawn in the last section.

\subsection{Einstein's equation and cosmological backgrounds}
To set notations and conventions, let us clarify that our aim is to consider spacetimes whose metric is
used in the description of the Universe. Hence, we stick to the standard convention of requiring the
Cosmological Principle to hold true; this straightforwardly leads to the full class of
Friedmann-Robertson-Walker metrics and, particularly, here we shall only consider those with a spatial flat 
section. In a Cartesian reference frame the metric reads
\begin{equation}\label{metric}
ds^2=-dt^2 + a(t)^2\delta_{ij}dx^idx^j,\quad i,j=1,...,3
\end{equation}
where $a(t)$ can be interpreted as usual as the expansion factor and it is the only function to be determined
out of (the semiclassical) Einstein's equations. A standard calculation shows that these can be reduced 
to an identity at a level of traces {\it i.e.} 
\begin{equation}\label{traceequation}
-R=8\pi\langle  T \rangle_\omega,
\end{equation}
together with the conservation law for the stress-energy tensor, namely
\begin{equation}\label{conservequation}
\nabla^a \langle T_{ab}\rangle_\omega=0.
\end{equation}
As already remarked in the introduction, $\langle T\rangle_\omega$ stands for the
expectation value of the stress-energy tensor. We shall deal with this issue more in detail in the
forthcoming discussion. As a last remark we wish to recall that \eqref{traceequation} and 
\eqref{conservequation} are actually not equivalent to a single but only to a set of Einstein's equations
which differ only by a conserved traceless tensor $T^0_{ab}$; such arbitrariness is fixed upon imposing  
suitable initial conditions.

\section{Massive scalar field.}

As we already emphasised in the introduction, we shall employ a real scalar field $\phi$ as the prototype to 
discuss the quantum behaviour of classical matter on a FRW background \eqref{metric}. Therefore the classical
dynamic of our system is governed by 
\begin{equation}\label{eqofm}
P\phi=0, \qquad P:= -\Box +\xi R + m^2, 
\end{equation}
where $\xi\in\mathbb{R}$, $R$ is the scalar curvature whereas $m$ is the mass of the field. Bearing in mind
that, unless stated otherwise, our convention for the metric signature is $(-,+,+,+)$, \eqref{metric} entails
the following identity $R=6\left( \frac{\ddot{a}}{a}+\frac{\dot{a}^2}{a^2} \right)$ where each dot stands for  
derivation with respect to $t$. In the next we shall indicate $H={\dot{a}}/{a}$. Setting $\xi=\frac{1}{6}$ 
corresponds to the so-called conformal coupling.

\subsection{Quantisation procedure: States and Hadamard condition}
In this paragraph we shall start dealing with the quantum behaviour of the solutions of \eqref{eqofm} and, to 
this avail, we
shall stick to the realm of the algebraic formulation of quantum field theory. Since a detailed analysis of
the main ingredients and results would require a review on its own just for the massive scalar field, we shall
point an interested reader to \cite{Haag,Wald}. Therefore, to cut a long story short, let us state that, to 
our purposes, it suffices to remember, that being the FRW spacetime, globally hyperbolic, it exists a 
standard procedure to
assign a $*-$algebra, say $\mathcal{W}$, out of \eqref{eqofm} \cite{Haag,Wald}. Afterwards we need to add a further 
ingredient, namely a state $\omega:\mathcal{W}\to\mathbb{C}$, which is the key tool out of which we can
calculate the relevant objects {\it i.e.} expectation values of the fields on that state, more commonly
referred to as $n-$point functions which we shall denote from now on as $\omega_n=\langle \phi(x_1)\dots \phi(x_
n) \rangle$. From a formal perspective these objects must be thought as distributions in $\mathcal{D}'(M^n)$ and the
singular structure, proper in general of distributions, arises whenever we perform in $\omega_n$ a coincidence 
limit.

Therefore, in order either to formulate a mathematically meaningful field theory either to construct a theory
which allows us to perform calculations going beyond the pure formal level, the selection of a suitable class
of states is one of the main, if not the most important, task. To this avail, we shall impose some reasonable 
constraints and the first requires us to restrict the attention to the so-called quasi-free states.
These are characterised by the following property: all the odd $n-$points functions vanish while all the even 
can be reconstructed out of sums of products of the two-points function. In other words quasi free states are
fully determined once $\omega_2(x,y)$ is known.
In the forthcoming sections we shall display how the above requirement is relevant to our discussion. In
particular we shall show that also the stress-energy tensor can be fully determined only out of $\omega_2$ and this is
the key non-geometrical ingredient in the semi-classical Einstein's equation.

Nonetheless ``quasi-free'' is not a sufficient requirement for our $\omega$ to satisfy and, particularly, a 
second and most important hypothesis must be imposed, namely the state shall be Hadamard. 
On a practical ground, from such a condition we can infer that the singular structure for the two-points
function is fixed as
\begin{equation}\label{two-points}
\omega_2(x,y)= \frac{1}{8\pi^2}\left(\frac{u(x,y)}{\sigma(x,y)}+v\log\sigma(x,y)+w(x,y)\right),
\end{equation}
where $\sigma$ is half of the square of the geodesic distance in the FRW background. The functions $u,v$ and 
$w$, also known as Hadamard coefficients, are smooth and $u$, $v$ can be uniquely determined once the 
equation of motion and the metric of the underlying background are fixed. In the above expression it turns 
out $u$ is the square root of the so-called van Vleeck-Morette determinant which depends only on $g_{ab}$,
{\it i.e.} $u$ can be reconstructed only out of the geometric properties of the manifold on which our fields
live. On the opposite, $w$ is the contribution to the Hadamard function which depends upon the state we
have selected.
Therefore all the information of the singular part in \eqref{two-points} is encoded in
$$
H(x,y)=\frac{1}{8\pi^2}\left(\frac{u(x,y)}{\sigma(x,y)}+v(x,y)\log\sigma(x,y)\right),
$$
which has a universal structure in every Hadamard state. Hence this is the contribution that we can subtract 
from the two-points function in order to get a smooth behaviour; in other words this amounts to regularise 
the state. As a notational convention, from now on, we shall refer to $v(x,x)$ by means of the symbol $[v]$. 
Furthermore $v(x,y)$ admits an asymptotic expansion in powers of the geodesic distance: 
$v(x,y)=\sum\limits_{n=0}^\infty v_n(x,y)\sigma^n(x,y)$. In 
the forthcoming discussion the coefficient $v_1$ will play a distinguished role.

\subsection{Stress-energy tensor}
The stress-energy tensor for a quantum real scalar field $\phi$ with mass $m$ and coupling to curvature $\xi$ 
can be written as
\begin{gather*}
T_{ab}:=\partial_a \phi \partial_b \phi - \frac{1}{6}g_{ab}\left( \partial_c\phi\partial^c\phi
+m^2\phi^2 \right)
-\xi \nabla_a\partial_b \phi^2+
\\
+\xi\left(  R_{ab}-\frac{R}{6} g_{ab} \right) \phi^2
+\left(\xi-\frac{1}{6}\right) g_{ab} \Box \phi^2.
\end{gather*}
Since the key ingredient to our analysis is the trace and the conservation equation for $T_{ab}$, let us
switch from the previous formula to
$$
T=-3\left(\frac{1}{6}-\xi \right) \Box \phi^2 -m^2 \phi^2, \qquad \nabla_a {T^a}_b=0.
$$
We stress to the reader that, here, we employ a non-standard form for $T_{ab}$, {\it i.e.} it differs from 
the more familiar one by a term proportional to 
$\frac{1}{3}\left( (P\phi)\phi +\phi (P \phi)\right) g_{ab}$ \cite{Mo03}. 
At a classical level this contribution vanishes since, on shell, $P\phi=0$, but nonetheless it represents an 
important feature in a full-fledged analysis of the underlying quantum theory, since, in this case, it is 
different from zero. Furthermore, encompassing such a term in the stress-energy tensor, automatically 
accounts for the trace anomaly which, on the opposite, was usually added by hand. As shown in 
\cite{Wald78,Fulling,HW01,HW02,Mo03}, this automatically arises in the quantum theory once the point splitting 
regularisation is performed. We also exploit the latter in order to regularise the operator $T_{ab}$ in 
order, subsequently, to calculate its expectation value on a quasi-free Hadamard state. Such an expression
would be quite cumbersome in the text and also of little avail; therefore an interested reader can refer to
the appendix A.1 for more details.

Notice that the envisaged conservation equation for the quantum stress-energy tensor, namely $\nabla_a\langle
{{{T}}^{ab}}\rangle_\omega = 0$, holds true due to the following identities 
$$
8\pi^2\langle \phi P\phi\rangle_{\omega} =6[v_1], \qquad 
8\pi^2\langle (\nabla_a\phi) (P\phi) \rangle_{\omega}=2\nabla_a[v_1], 
$$
where $[v_1]$ is here explicitly given in the appendix in formula \eqref{v_1}.
The heritage of such a conservation law is the change of the expectation value for the trace of $T_{ab}$ by 
means of a purely quantum term: 
$$
\langle {T}\rangle_{\omega}:=\left(-3\left(\frac{1}{6}-\xi\right)\Box -m^2\right) \frac{[w]}{8\pi^2}+\frac{2[v_1]}
{8\pi^2},
$$
where the dependence upon the state is encoded in the term $[w]$.

To conclude, we point out to a potential reader that, due to $[v_1]$, the above trace is non vanishing also 
in a conformal field theory \cite{Wald78}.

\subsection{Remaining freedom in the definition of $T_{ab}$}
By means of point splitting regularisation we have fixed the expectation value of $\langle T\rangle_\omega$ in 
the so-called minimal regularisation prescription, namely we have only subtracted the singular part form the 
two-points function. 
Nonetheless, as discussed by Wald \cite{Wald78}, in the renormalisation prescription, there is still a 
freedom of geometric nature. In detail we can add a tensor $t_{ab}$ written only in term of the local metric
and such that it satisfies $\nabla^at_{ab}=0$ without either affecting the equations of motion for the matter 
either violating the first four axioms introduced and discussed in Wald paper. 
The conservation equation for $t_{ab}$ is not the unique constraint we may wish to impose on such a tensor
and, in particular, a further natural requirement would be that $t_{ab}$ behaves as $T_{ab}$ under scale 
transformations. 
In other words this implies that $t_{ab}$ arises out of the following variation 
$$
t_{ab}=\frac{\delta}{\delta g^{ab}}\int{ A\sqrt{g} R^2 + B\sqrt{g}R_{ab}R^{ab}  },
$$ 
being $A$ and $B$ just arbitrary real numbers.
Leaving the details of the above construction and analysis to \cite{Wald78, HW01,HW02}, we shall only 
stress that the trace of $t_{ab}$ turns out to be proportional to $\Box R$ independently from the choice of 
$A$ and $B$. This is an unavoidable arbitrariness in the employed scheme and, as a byproduct, it leads us to think of $A$ 
and $B$ as renormalisation constants on their own. 
We are now able to compute the trace of the whole quantum modified stress-energy tensor:
$$
\langle {T}\rangle_{\omega}:=\left(-3\left(\frac{1}{6}-\xi\right)\Box -m^2\right) \frac{\langle\phi^2\rangle_\omega}{8\pi^2}+\frac{2[v_1]}{8\pi^2}
+c\Box R,
$$
where $c$ is a linear combination of $A$ and $B$ and it represents the freedom in the renormalisation 
procedure we exploited. Eventually, $c$ will be chosen in order for the trace to satisfy the requirement
coming out of the fifth Wald's axiom (still see \cite{Wald78}); in other words there must be no derivatives of the 
metric with degree higher than 2 in the expectation values of ${T}_{ab}$.
The remaining renormalisation ambiguity is encoded in the expectation value of the filed 
$\langle\phi^2\rangle_\omega$; we shall come back later to this point fixing the ambiguity by physical 
motivation.

We stress that a similar observation brought interest in the so-called modified theory of gravity also known 
as $f(R)$ gravity. Nonetheless the view we wish to push home is the following: adding $t_{ab}$ does not come 
from a modified gravitational action, but it only originates form the employed renormalisation scheme, {\it
i.e.} it must be an effect coming from quantum matter.
Naturally this does not exclude that such a perspective cannot provide hints on how a candidate theory of 
quantum gravity interacts with quantum matter.
As a final comment we would like to stress that the above is the subtlest point in the whole construction. We 
used an expression for the stress-energy tensor which is suitable in order to deal with semiclassical Einstein's 
equation. Nonetheless such a modification is not artificial, corresponding as a matter of facts just to a 
specific choice  of the renormalisation constants arising out of the employed scheme. 

\section{Evolution equation of the model}
In the case of conformal coupling $\xi=1/6$, equation \eqref{traceequation}, written in terms of 
$H=\dot{a}/a$, becomes
\begin{gather}
-6\left(\dot{H}+2 H^2\right)=
-8\pi G m^2 \langle\phi^2\rangle_\omega+
\notag\\ \label{eqdiff}
+\frac{G}{\pi}\left(
-\frac{1}{30}\left(\dot{H}H^2+H^4 \right) 
+\frac{m^4}{4}\right).
\end{gather}
The aim of this section is to analyse in detail the possible solutions of \eqref{eqdiff} under some specific
hypotheses on the expectation value for $\langle\phi^2\rangle_\omega$. Particularly we shall show that 
a de Sitter space with a specific curvature will appear as  a stable solution.

\subsection{Conformal invariant case: stability of de Sitter phase}
As a starting point we shall deal with the scenario in which $m=0$, already encompassed in 
Starobinsky paper \cite{Star80} (see also \cite{Vile85}). As remarked above, there is no need to select a
specific state and an ordinary differential equation rules the evolution of $H$. 
Hence, setting $m=0$ in \eqref{eqdiff}, we end up with 
\begin{equation}\label{eqc}
\dot{H}\left( H^2-H_0^2\right)=-H^4+2H_0^2H^2.
\end{equation}
Here $H_0^2=\frac{180\pi}{G}$ depends on the Newton 
constant and it has an order of magnitude of 24 times the inverse Planck time. Let us notice that, out of the
right hand side of \eqref{eqc}, we can extract
two critical points; therefore \eqref{eqc} admits two constant solutions, namely $H(t)=0$ and
$H(t)=H_+=\sqrt{2} H_0$ corresponding respectively to a Minkowski spacetime and to a de Sitter one.
Suppose now to assign an initial condition at a fixed time $t_0$ such that $H(t_0)\neq 0$ and $H(t_0)\neq
H_+$; we are interested to realize if the solution interpolating such an initial condition flows at large
times either to $0$ or to $H_+$ {\it i.e.}, in order words, whether these two critical points are stable or
not. To bring such task to a good end, we simply need to notice that \eqref{eqc} is integrable as:
\begin{equation}\label{sol}
K e^{4t}=e^{{2}/{H}}\left|\frac{H+H_+}{H-H_+}\right|^{1/H_+},
\end{equation}
where $K$ stands for the integration constant to be fixed out of the initial condition $H_0$. Depending on
such last value, all the solutions $H(t)$ flow either to $0$ or to $H_+$. Hence both critical points turn out 
to be stable. This result is different from the classical outcome of the analysis due to Starobinsky 
\cite{Star80} (see also Vilenkin and Ford \cite{Vile85, Ford84}). The price to pay, in order to achieve such 
a result, is a choice by hand of a renormalisation constant. It turns out to be an addition of a tensor 
written only in terms of the metric and such operation introduces in the theory a scale-length, as already 
discussed by Wald in \cite{Wald78}. We have to stress that, on the dark side, the above de Sitter solution 
cannot describe the present days form of the universe  being $H_+\simeq 6.4 \times 10^{44} s^{-1}$ {\it i.e.}
many orders of magnitude bigger then the present measured Hubble constant $(2.6\pm 0.2) \times 10^{-18} s^{-1
}$. On the bright side, instead, we have shown that, encompassing the full quantum effects, we are lead to 
find a stable de Sitter solution even if no cosmological constant is present in the equations.

\subsection{Massive case with $\xi=1/6$: stability of the de Sitter phase, effective cosmological constant} 
\begin{figure}
  \centering
\includegraphics[width=\columnwidth]{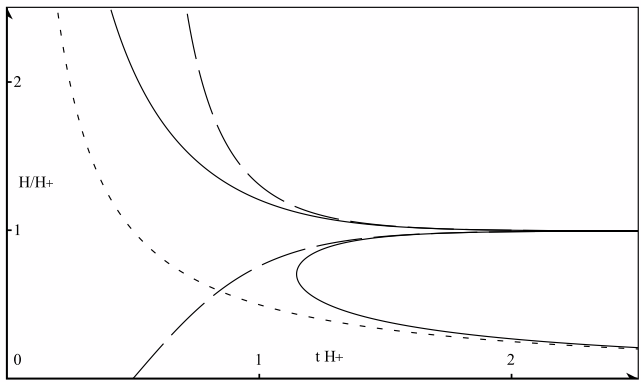}
\caption{\small Here the dashed line corresponds to the behaviour of $\frac{H}{H_+}$ as a function of time
$t$ (normalised with respect to $1/H_+$) in a FRW universe with a non vanishing cosmological constant and filled with radiation, while the dotted 
line stands for the lone classical contribution coming from radiation. Conversely the continuous line depicts 
the outcome of our model where quantum effects are also taken into account.}\label{unica}
  \label{soluzione}
\end{figure}

In this section we switch from the massless to the massive case. 
The most important difference is the following: the righthand side of \eqref{eqdiff} depends explicitly upon 
the state via the expectation value of $\phi^2$.
The expectation value of $\langle\phi^2\rangle_\omega$ on a general Hadamard state $\omega$ is
$\frac{[w]}{8\pi^2}+\alpha m^2 + \beta R$,
where $\alpha$ and $\beta$ are renormalisation constants encoding the ambiguities still present in the 
procedure. We assume for the moment the existence of a set of Hadamard states $\widetilde\omega$, one for each 
spacetime whose metric is of the form \eqref{metric} being $H=\dot{a}/a$ and $\langle\phi^2\rangle_{\widetilde
\omega}=\alpha m^2 + \beta R$. We shall see later that this 
assumption turns out to be an approximation of the expectation values of the fields computed on the adiabatic
states of FRW in the limit where $m^2>>R$ and $m>>H$. Moreover, by the principle of general local covariance 
\cite{HW01,HW02,BFV}, we are entitled to fix the renormalisation constants once and in the same way for every
spacetime we are considering. Then the expectation value of $\langle\phi^2\rangle_{\widetilde\omega}$ on the 
states we are 
considering takes the following values:
\begin{equation}\label{2pt}
\langle\phi^2\rangle_{\widetilde\omega}=\alpha m^2 + \beta R,
\end{equation}
on all the considered FRW spacetimes.
Therefore, taking into account these remarks, \eqref{eqdiff} takes the following form:
\begin{equation}\label{massiveeqdiff}
\dot{H}\left( H^2-H_0^2\right)=-H^4+2H_0^2H^2+M,
\end{equation}
where $H_0$ and $M$ are the following two constants with the following values
$$
H_0^2=\frac{180\pi}{G}-8\pi^2 180 m^2\beta, 
\qquad M=\frac{15}{2}m^4-240\pi^2 m^4 \alpha.
$$
As in the previous section, the right hand side of \eqref{massiveeqdiff} displays at most two critical
points amounting to
\begin{equation}\label{fixedp}
H_{\pm}^2=H_0^2\pm\sqrt{H_0^4+M},
\end{equation}
both corresponding either to a de Sitter phase or to a Minkowski phase. 
A straightforward analysis shows that both $H(t)=H_\pm$ appear to be stable since all the solutions flow to 
either one of the two fixed points. It is remarkable that the existence and the stability behaviour of the 
latter is left unchanged whether the right hand side of \eqref{2pt} is modified adding a term such as 
$Aa^{-\lambda}(t)$, being $\lambda\in\mathbb{R}$ and $A$ a constant of suitable dimension. 
It is also interesting to notice that a formula similar to \eqref{fixedp} already appeared in \cite{pst} although, in the cited paper, a classical cosmological constant has been introduced from the beginning.
At this stage our 
simple model depends on three parameters $\alpha,\beta,
m$. A minimal and, to a certain extent, compulsory choice is to require Minkowski as a solution of our 
system. This amounts to fix $\alpha=(32\pi^2)^{-1}$ which, on the other hand, entails $M=0$. The form of the 
solution is then equal to that of the massless case \eqref{sol}, where one of the fixed points corresponds to 
a Minkowski space - $H(t)=0$ -, while the other fixed point $H(t)=H_+$ corresponds to de Sitter.
With respect to the massless conformal factor, here we can fine-tune the parameters $\beta$ and $m$ in such a
way for $H_+$ to be small enough in order to account for the present measured value of the Hubble constant. 
Hence, heuristically speaking, our system behaves as if an effective cosmological constant enters the fray
without even being present at the beginning and this is a strict consequence of encompassing the full quantum 
properties of the field. 
As a further remark we would like to notice that \eqref{sol} displays, for a large class of initial 
conditions, an early time phase of rapid expansion which is a prerequisite feature of modern
models for studying the early stages of evolution of the Universe. This is in sharp contrast with the canonical
paradigm according to which quantum effects should account only for small fluctuations with respect to the
classical behaviour. On the opposite, even in the most simple example of a massive scalar field and with the
most simple assumptions, our system displays a behaviour which drastically differs from the one we could a
priori expect only from a classical analysis. Hence this suggests that, when dealing with scalar fields on a 
FRW background, one should always perform a full-fledged analysis of the semiclassical behaviour of the 
system since the quantum contributions appear to be hardly negligible as one can also infer from figure
\ref{unica}.

As a final comment we would like to stress that, in a neighbourhood of  $H=H_+$, the found solution 
\eqref{sol} looks rather 
similar to the one of a classical flat universe with cosmological constant filled with radiation.
As a matter of fact, in that case $H(t)=A\tanh(2(t-t_0) A)$ where $A$ is a constant related to the 
cosmological constant, and it can be inverted as 
$$
K e^{4t}=\left|\frac{H+A}{H-A}\right|^{1/A},
$$
which looks very similar to \eqref{sol} when $H\sim A$ and $H_+=A$; this corresponds to the dashed line in  
figure \ref{unica}. The quantum effects are not important only around $H=0$ where \eqref{sol} looks like  
$H(t)$ in a flat universe filled only with radiation, namely the dotted line in the figure \ref{unica}. 
Eventually we would like to stress 
that considering the upper brunch of the solution, in the past, it displays the behaviour of a classical flat 
universe with a kind of matter such that $\rho=A\; a(t)^{-2}$. Even in this regime quantum effects are not 
negligible. As a further remarkable consequence of the analytic form of $H(t)$, it turns out that 
the singularity at $t=t_0$ coincides with null past infinity in the flat spacetime conformally related to 
\eqref{metric}; hence it descends that the particle horizon is not present. Therefore any pair of points in 
the underlying background was casually related in the past, and, thus, as a byproduct, such property of our 
model could provide a solution to the problem of homogeneity.

\section{Expectation value of $\phi^2$ on the adiabatic vacuum}

In the preceding section we have seen that, assuming a suitable form of $\langle\phi^2\rangle$, two stable de 
Sitter phases can arise as solutions of the semiclassical Einstein's equation.
We would like to give a justification for our assumption, namely we shall show that there is a regime in
which it is is valid.
Here we restrict our attention to the case of a massive scalar field with a conformal coupling to the 
metric. The first observation is that, if we select the Bunch-Davies state $\omega_B$ \cite{BD} on a de Sitter 
spacetime and if we compute the renormalised version of the expectation value of $\phi^2$, we obtain a 
constant that depends only on the mass $m$ and on $H$.
With this observation we can immediately conclude that the two fixed points $H(t)=H_+$ and $H(t)=H_-$ 
discussed above are really exact solutions of the semiclassical Einstein's equation.
In the next we shall select a class of states that, in the limit of a large mass, shows an expectation value 
for $\langle \phi^2\rangle_\omega$ that is of the type $\alpha m^2 + \beta R$.

\subsection{Adiabatic states and large mass expansion}
We would like to select here the class of adiabatic states, {\it i.e.} those introduced by Parker 
\cite{Parker} in order to minimise particle creation (see also \cite{ZeSt} for a derivation of the 
expectation values of the stress tensor). Much work has been done also recently in order to make 
the definition of these states precise \cite{LuRo,JuSc,Olbermann}. 
In order to write the two-points function of these states we follow the construction as in Parker 
\cite{Parker}. In the case of conformal coupling it is convenient to use the conformal time $\tau$ defined as
 $\tau-\tau_0=\int_{t_0}^t \frac{dt'}{a(t)}$. Therefore the two-points function of such kind of states is
$$
\omega(x_1,x_2)=\frac{1}{8\pi^3}\frac{1}{a(\tau_1)a(\tau_2)}\int d^3 {\bf k} \overline{\Psi_k(\tau_1)} \Psi_k(
\tau_2) e^{i {\bf k}\cdot{({\bf x}_1-{\bf x}_2)}};
$$
above $x_i$ $k_i$ are four vectors and  ${\bf x}_i$ are three vectors whereas $|{\bf k}|$ stands for the 
length of the spatial vector ${\bf k}$.
The functions $\Psi_k(\tau)$ are solutions of a differential equation with a suitable normalisation 
condition:
\begin{gather*}
\left(\frac{d^2}{d\tau^2} +k^2 +m^2 a(\tau)^2 \right) \Psi_k(\tau)=0,\\\overline{\Psi_k(\tau)}\frac{d}{d\tau}\Psi_k(\tau)-\Psi_k(\tau)
\frac{d}{d\tau}\overline{\Psi_k(\tau)} = i.
\end{gather*}
Each $\Psi_k(\tau)$ can alternatively be written in the following way:
$$
\Psi_k(\tau)=\frac{1}{\sqrt{2 \Omega_k(\tau)}} e^{i\int_{\tau_0}^\tau \Omega_k(\tau)}.
$$
In the adiabatic approximation  $\Omega_k(\tau)$ is a function constructed recursively in the following way: 
$${\Omega_k^{(0)}}^2(\tau)=k^2+m^2 a(\tau)^2,$$ 
and 
\begin{equation}\label{ricorsione}
{\Omega^{(n+1)}_k}^2 (\tau)=  k^2+m^2 a(\tau)^2+\frac{3}{4}\left( \frac{{\Omega_k^{(n)}}'(\tau)}{\Omega_k^{(n)}(\tau)}\right)^2-
\frac{1}{2}\frac{{\Omega_k^{(n)}}''(\tau)}{\Omega_k^{(n)}(\tau)},
\end{equation}
where the prime stands for the derivation with respect to $\tau$.
The $n$-th order approximation consists then in the substitution of $\Omega_k$ with $\Omega_k^{(n)}$ in 
$\Psi_k(\tau)$ and we shall indicate with $\omega^{(n)}_2$ the counterpart for the two-points function of the 
state. Nonetheless one should bear in mind that this recursive procedure does not have nice convergence 
properties though, thanks to the work of Junker and Schrohe \cite{JuSc}, we know that the state constructed in
this way is an adiabatic state in the sense that $\omega^{(n)}_2$ have a certain Sobolev wavefront set.
Hence, if $n$ is large enough, we can use the approximated state in order to build the stress-energy tensor 
or the expectation value of $\phi^2$.
In particular, we can compute the approximated expectation value
$\langle{\phi}^2\rangle_{(n)}=\lim\limits_{x\to y}(\omega_{2}^{(n)}(x,y)-H(x,y))$, 
which, more explicitly, becomes
\begin{gather*}
\langle{\phi}^2\rangle_{(n)}=\frac{1}{{4\pi^2}\;a(\tau)^2}\int_0^\infty dk\; k^2 \left(\frac{1}{\Omega^{(n)}_k(\tau)}-\frac{1}{\Omega_k^{(0)}(\tau)} \right)+\\
+\alpha' R +\beta' m^2.
\end{gather*}
Above $\alpha'$ and $\beta'$ need to be interpreted as renormalisation constants.
An exact computation of this integral can be very difficult to perform, hence we will show only how to 
compute an expectation value in the limit of a large mass, namely, assuming that $H(t)$ is a smooth function 
and $m^2>>R$. 
In this case, if furthermore $n\geq2$, it is possible to expand the integral in  powers of $1/m^2$, as:
$$
\langle{\phi}^2\rangle_{(n)}=\alpha m^2 + \beta R+O\left(\frac{1}{m^2}\right),
$$ 
where $\alpha$ and $\beta$ are slightly different from the one written before.
In the large mass limit we shall simply consider $\langle{\phi}^2\rangle_{(n)}=\alpha m^2 + \beta R$.
The result should be read as a confirmation for the approximation we have done in the preceding section.

\section{Interpretation of the results and final comments}

In the present paper we have shown that, when dealing with cosmological models, quantum effects are not 
negligible even when we consider basic models. As a matter of facts, our analysis displays that, from a 
careful analysis of the expectation values of the renormalised stress-energy tensor, it arises an 
effective cosmological constant which can be interpreted as dark energy.

Such a feature is manifest if we take into account a massive scalar field propagating in a curved background,
although we envisage that similar effects would be present if we consider other kinds of fields.
Furthermore we have seen that a de Sitter solution appears as a stable fixed point of the semiclassical 
Einstein's equation and, to a certain extent, also a phase of rapid expansion can be foreseen in the model.
We also believe that, since the found results, and particularly the stability of the de Sitter solution, are
based upon a modification of the point splitting procedure by a pure gravitational term, this could be read 
as an hint for future study of quantum gravitational models interacting with matter. To this avail it also 
seems interesting to pinpoint that, even considering the one-loop corrections to the action of an $f(R)$ 
theory, one is lead to a stable of de Sitter solution \cite{CZ05, CZ06}. Furthermore, also in this last case,
stability is a joint effect of quantum theory and classical gravity and this is a behaviour which a lone 
$f(R)=R^2$ term does not display.

\section*{Acknowledgements.} The work of C.D. is supported by the von Humboldt Foundation and that of N.P. 
has been supported by the German DFG Research Program SFB 676. We would like to thank 
R. Brunetti, S. Hollands, V. Moretti and R. M. Wald for useful discussions. 
We are also grateful to I. L. Shapiro and A. A. Starobinsky for useful comments and remarks. 
 
\appendix
\section{Point splitting regularisation of the stress-energy tensor}
Let $\omega_2$ be the two-points function of a quasi free Hadamard state. The expectation value of the
stress-energy tensor regularised by means of the point splitting procedure is: 
\begin{gather*}
\langle T_{ab}\rangle_{\omega}(z):=
\lim_{(y,x)\to (z,z)} 
\left[\partial_a\partial'_b- \frac{1}{6}g_{ab}\left( g^{cd}\partial_d{\partial'}_c
+m^2\right)\right. +\\
-2\xi \left(\nabla_a\partial_b+\partial_a\partial_b' \right) +\xi \left(R_{ab}(z)-\frac{R(z)}{6} g_{ab}\right)+\\
\left.+(\xi-\frac{1}{6}) g_{ab} (2\nabla^c\partial_c+2g^{dc}(z)\partial_d \partial'_c)\right]\\
\frac{1}{2}\left(\omega_2(y,x)-H(y,x) + \omega_2(x,y)-H(x,y) \right).
\end{gather*}
where the prime stands for a derivative in $y$ whereas the one without prime is a derivative with respect to 
$x$. A reader should notice that, in the last part of the equation, there is a symmetrisation done at the 
level of two-points function and that $H(x,y)$ is the singular part of the Hadamard series.

\subsection{$[v_1]$ coefficient in the cosmological case}
Since it is a relevant datum in our procedure, we provide the explicit expression for $2[v_1]=[a_2]/2$, being 
$a_2$  the Schwinger-de Witt coefficient as derived at pag. 194 in \cite{Fulling} with the choice of $V=\xi R+
m^2$, (see also \cite{Tadaki})
\begin{gather}\notag
2[v_1]=\frac{1}{360}\left( C_{ijkl}C^{ijkl}+R_{ij}R^{ij}-\frac{R^2}{3}+\Box R\right)+\\
+\frac{1}{4}\left(\frac{1}{6}-\xi\right)^2R^2
+\frac{m^4}{4}
-\frac{1}{2}\left(\frac{1}{6}-\xi\right){m^2}R+\notag\\
+\frac{1}{12}\left(\frac{1}{6}-\xi\right)\Box R.\label{v_1} 
\end{gather}
Furthermore, assuming that the metric has the form of a flat FRW universe \eqref{metric} and 
writing $H=\dot{a}/a$, $[v_1]$ takes the following form
\begin{gather*}
2[v_1]=-\frac{1}{30}\left(\dot{H}H^2+H^4 \right) 
+\frac{1}{12}\left( \frac{1}{5}-\xi\right) \Box R+ \notag\\
+9\left(\frac{1}{6}-\xi\right)^2
\left(\dot{H}^2+4H^2\dot{H}+4 H^4 \right)
+\frac{m^4}{4}+\\
-3\left(\frac{1}{6}-\xi\right){m^2}\left(\dot{H}+2 H^2\right).
\end{gather*}

\bibliography{DFP_PRD}

\end{document}